\documentstyle[12pt,epsf]{article}

\def\lesssim{\mathrel{\mathpalette\vereq<}}

\makeatletter
\def\vereq#1#2{\lower3pt\vbox{\baselineskip1.5pt \lineskip1.5pt
\ialign{$\m@th#1\hfill##\hfil$\crcr#2\crcr\sim\crcr}}}
\makeatother

\DeclareRobustCommand\verylongrightarrow
     {\relbar\joinrel\relbar\joinrel\relbar\joinrel\relbar\joinrel\rightarrow}
\begin{document}

%%%%%%%%%%%%%%%%%%%%%%%%%%%%%%%%%%%%%%%%%%%%%%%%%%%%%%%%%%%%%%
\catcode`@=11
% Redefine caption to put text and formulas in smaller font
\long\def\@caption#1[#2]#3{\par\addcontentsline{\csname
  ext@#1\endcsname}{#1}{\protect\numberline{\csname
  the#1\endcsname}{\ignorespaces #2}}\begingroup \small
  \@parboxrestore \@makecaption{\csname
  fnum@#1\endcsname}{\ignorespaces #3}\par \endgroup} 
\catcode`@=12
%%%%%%%%%%%%%%%%%%%%%%%%%%%%%%%%%%%%%%%%%%%%%%%%%%%%%%%%%%%%
\newcommand{\newc}{\newcommand}
\newc{\gsim}{\lower.7ex\hbox{$\;\stackrel{\textstyle>}{\sim}\;$}}
\newc{\lsim}{\lower.7ex\hbox{$\;\stackrel{\textstyle<}{\sim}\;$}}
\newc{\gev}{\,{\rm GeV}} 
\newc{\mev}{\,{\rm MeV}} 
\newc{\ev}{\,{\rm eV}} 
\newc{\kev}{\,{\rm keV}} 
\newc{\tev}{\,{\rm TeV}} 
\newc{\mz}{m_Z}
\newc{\mpl}{M_{Pl}}
\newc\order{{\cal O}} 
\newc\CO{\order} 
\newc\CL{{\cal L}}
\newc\CM{{\cal M}}
\newc{\eps}{\epsilon} 
\newc{\re}{\mbox{Re}\,} 
\newc{\im}{\mbox{Im}\,}
\newc{\invpb}{\,\mbox{pb}^{-1}} 
\newc{\invfb}{\,\mbox{fb}^{-1}}
\newc{\str}{\mbox{STr}\,} 
\newc{\tr}{\mbox{Tr}\,} 
\newc{\mh}{m_h}
\newc{\mt}{m_t} 
\newc{\mw}{m_W} 
\def\vev#1{\langle #1 \rangle}
\newc{\Lambdamax}{\Lambda_{max}} 
\newc{\CF}{{\cal F}}
\newc{\mgb}{m_{\mbox{\scriptsize NGB}}}
\newc{\msusy}{M_{\mbox{\scriptsize SUSY}}}
%%%%%%%%%%%%%%%%%% Reference Defs %%%%%%%%%%%%%%%%%%
%
\def\NPB#1#2#3{Nucl. Phys. {\bf B#1} (19#2) #3}
\def\PLB#1#2#3{Phys. Lett. {\bf B#1} (19#2) #3}
\def\PLBold#1#2#3{Phys. Lett. {\bf#1B} (19#2) #3}
\def\PRD#1#2#3{Phys. Rev. {\bf D#1} (19#2) #3}
\def\PRL#1#2#3{Phys. Rev. Lett. {\bf#1} (19#2) #3}
\def\PRT#1#2#3{Phys. Rep. {\bf#1} (19#2) #3}
\def\ARAA#1#2#3{Ann. Rev. Astron. Astrophys. {\bf#1} (19#2) #3}
\def\ARNP#1#2#3{Ann. Rev. Nucl. Part. Sci. {\bf#1} (19#2) #3}
\def\MPL#1#2#3{Mod. Phys. Lett. {\bf #1} (19#2) #3}
\def\ZPC#1#2#3{Zeit. f\"ur Physik {\bf C#1} (19#2) #3}
\def\APJ#1#2#3{Ap. J. {\bf #1} (19#2) #3} 
\def\AP#1#2#3{{Ann. Phys. }{\bf #1} (19#2) #3} 
\def\RMP#1#2#3{{Rev. Mod. Phys. } {\bf #1} (19#2) #3} 
\def\CMP#1#2#3{{Comm. Math. Phys. } {\bf #1} (19#2) #3} \relax
%
%
%%%%%%%%%%%%%%%%%%%%%%% latex eqn abrev's %%%%%%%%%%%%%%%%%%%%%%%%%%%%
%
\def\beq{\begin{equation}} 
\def\eeq{\end{equation}}
\def\bea{\begin{eqnarray}} 
\def\eea{\end{eqnarray}}
%
%%%%%%%%%%%%%%%%%%%%%%% common abrev's %%%%%%%%%%%%%%%%%
%
%
\newc{\ie}{{\it i.e.}}  
\newc{\etal}{{\it et al.}}  
\newc{\eg}{{\it e.g.}}  
\newc{\etc}{{\it etc.}}  
\newc{\cf}{{\it c.f.}}
%
%
%%%%%%%%%%%%%%%%%%%% slashed symbols %%%%%%%%%%%%%%%%%%%%%
%
%
\def\slash#1{\rlap{$#1$}/} % slashes a character
\def\Dsl{\,\raise.15ex\hbox{/}\mkern-13.5mu D} %this one can be subscripted 
\def\delsl{\raise.15ex\hbox{/}\kern-.57em\partial}
\def\Ksl{\hbox{/\kern-.6000em\rm K}} 
\def\Asl{\hbox{/\kern-.6500em \rm A}}
\def\Qsl{\hbox{/\kern-.6000em\rm Q}}
\def\gradsl{\hbox{/\kern-.6500em$\nabla$}}
%
%%%%%%%%%%%%%%%%%%% various symbol abbreviations, vev's etc %%%%%%%%%%%
%
%
\def\bar#1{\overline{#1}} 
\def\vev#1{\left\langle #1 \right\rangle}
%
%%%%%%%%%%%%%%%%%%% end of intro %%%%%%%%%%%%%%%%%%%%%%%%%%%%%%%%%%%%%

\begin{titlepage}
\begin{flushright}
{LBNL-45295\\ UCB-PTH-00/09\\ hep-ph/0003170\\ March 2000}
\end{flushright}
\vskip 2cm
\begin{center}
%{\large\bf Naturalness}
{\large\bf The Higgs Mass and New Physics Scales \\
in the Minimal Standard Model\footnote{This work was supported in
part by the Department of Energy under contract DE--AC03--76SF00098
and in part by the National Science Foundation
under grant PHY-95-14797.}}
\vskip 1cm {\normalsize Christopher Kolda and Hitoshi Murayama}\\
\vskip 0.5cm {\it Department of Physics\\ University of California \\
Berkeley, CA~~94720, USA\\ and \\ Theory Group\\ Lawrence Berkeley
National Laboratory\\ Berkeley, CA~~94720, USA\\}
\end{center}
\vskip .5cm
\begin{abstract}
  We study the theoretical correlation between the Higgs mass of the
  minimal standard model and the scale at which new physics is
  expected to occur. In addition to the
  classic constraints of unitarity, triviality and vacuum stability,
  we reexamine the constraints imposed by the precision electroweak
  data. We then pay particular attention to the constraint imposed by 
  the absence of fine-tuning in the Higgs mass parameter (the Veltman
  condition).  We find that the fine-tuning condition places a
  significant constraint on the new physics scale for the Higgs mass
  range $100\gev \lesssim m_{h} \lesssim 200\gev$ mostly unconstrained by the
  classic constraints.
\end{abstract}
\end{titlepage}
\setcounter{footnote}{0} 
\setcounter{page}{1} 
\setcounter{section}{0}
\setcounter{subsection}{0} 
\setcounter{subsubsection}{0}

%%%%%%%%%%%%%%%%%%%%%%%%%%%%%%%%%%%%%%%%%%%%%%%%%%%%%%%%%%%%%%%%%%%%%%%

\section{Introduction} \label{sec:intro}

The Standard Model of particle physics has passed every experimental
challenge to date, the only important exception being neutrino
oscillations.  The gauge sector, namely the interactions of quarks and
leptons with gauge bosons, has been verified at the {\it permille}\/
level due mostly to the LEP and SLC experiments~\cite{LEPEWWG}.  The
Yukawa sector is being tested vigorously by exploiting flavor-changing
and CP-violating phenomena.  The last year has produced two new
manifestations of CP violation~\cite{sin2b,epsilon'} a quarter century
after the first evidence, and higher luminosity $B$-experiments will
explore this sector even further.  Yet the most mysterious and the
least understood sector is the electroweak symmetry breaking sector.
In the minimal standard model, our lack of understanding is
parameterized by the mass of the Higgs boson, $m_{h}$.

There is actually another unknown parameter of the standard model. It
is commonly assumed that the minimal standard model is merely a
low-energy effective theory of some more fundamental theory which
explains the origin of electroweak symmetry breaking. Such a theory
may also explain the structure of fermion masses and mixings, the 
strange assignments of fermion gauge charges, and so forth.
The energy scale, $\Lambda$,
of such a (more) fundamental theory provides an ultraviolet cutoff for
the standard model but is completely unknown at present.

In the absence of direct experimental information, theoretical
considerations are our only guide to constraining the ranges of
$m_{h}$ and $\Lambda$.  An important example of such an aid is the
electroweak precision analysis, where the precise measurements of
masses, widths and asymmetries indirectly constrain the Higgs mass and
new physics through loop diagrams.  In this letter, we extend such
analyses to determine the possible range of $m_{h}$ and $\Lambda$
based on various theoretical requirements. These include the
traditional analyses of unitarity~\cite{unitarity}, 
triviality~\cite{trivial} and vacuum stability~\cite{stability}.  
We will revisit
the precision electroweak analysis of Hall and Kolda~\cite{hk} using
improved methodology and newer data. And finally, we will impose
naturalness/fine-tuning constraints on the Higgs sector. We will find
that naturalness may not be as constraining as is generally believed,
but combined with the other requirements will provide a narrow region
of interest in the search for new physics. This will be important in
planning future high-energy experiments.

This type of discussion can be used in two different ways.  If one
picks a particular candidate for new physics with an associated energy
scale $\Lambda$, a certain range of Higgs mass is preferred; thus we
predict $m_{h}$ given a framework.  On the other hand, if we find and
measure the mass of the Higgs boson, this analysis provides an upper
bound on the new physics scale $\Lambda$.  Either way, the constraints
on $(m_{h}, \Lambda)$ will be valuable.

The new physics at scale $\Lambda$ should protect the smallness of the
electroweak scale $v = 175\gev$ against the corrections due to physics
at the Planck scale, $M_{Pl} = 10^{19}\,$GeV.  The most often
discussed candidate for such physics is probably supersymmetry (SUSY)
(see, {\it e.g.}\/, \cite{ICTP} for a review), but other possibilities
(\eg, a composite Higgs sector~\cite{technicolor,compohiggs}) also
exist.  An alternative possibility is that the Planck-scale is
actually much lower than apparent because of large extra 
dimensions~\cite{extraD}.  Our analysis avoids specifying such new physics
explicitly, and we try to remain as model-independent as possible.
This unfortunately introduces an uncertainty in the definition of the
scale $\Lambda$. However for most choices of new physics, $\Lambda$
represents roughly the scale of some particle which has been
integrated out to produce the standard model. This is of course the
scale most relevant to direct searches at high energy colliders.

This letter is organized as follows.  In the sections that follow, we
explain the theoretical requirements on $m_{h}$ and $\Lambda$,
starting with the well-known constraints coming from unitarity,
triviality and vacuum stability. We then extend these to include the
now-powerful arguments coming from the precision electroweak analysis.
Finally we revisit an old idea in new clothing, namely the absence of
fine-tuning and the so-called Veltman condition~\cite{veltman}.
Finally we comment on the special case of supersymmetry briefly.

\section{The Usual Suspects}

Unitarity, triviality and vacuum stability are the classic triumvirate
of Higgs mass constraints. Over the years, these three ideas have
repeatedly guided our thinking about Higgs physics and scales of new
physics, and are essential parts of any such discussion. But since
they are so well-known we will only give a short discussion of them
here.

\subsection{Unitarity}

Unitarity is simultaneously the strongest and weakest constraint when
considering the Higgs mass and new physics~\cite{unitarity}. It is the
strongest in that it provides an upper bound on $\mh$ above which the
standard Model is known to become non-perturbative. It is the weakest
because it provides only binary information about that new physics: if
$\mh$ exceeds its unitarity bound, then the new physics must appear
almost immediately; if $\mh$ is less than its unitarity bound, then
there is no need for new physics and there is nothing further to be
said.

In this context, ``unitarity'' should more properly be called
``tree-level unitarity'' because it is the requirement that the
tree-level contribution to the first partial wave in the expansion of
various scattering amplitudes not exceed the unitarity bound. This
breakdown of unitarity at the lowest order can either be cancelled by
some new physics which plays the role of the Higgs,
% (\ie, by providing the longitudinal degree of freedom for the weak
% gauge bosons), 
or by a cancellation of the tree-level against higher
orders. The first implies new physics {\it prima facie}. The second
implies a breakdown in perturbation theory at a nearby scale, which
we will consider to be in itself ``new physics'' for the purposes of
this article.

The calculation is simple and well-known (we follow the
second paper in Ref.~\cite{unitarity}), but helps to
define our notation. Begin with the tree-level Higgs potential:
\beq
V_0=\mu^2 |\phi|^2+\lambda |\phi|^4.  
\eeq 
This potential satisfies the well-known minimization condition
$|\phi|^2=-\mu^2/2\lambda$. Expanding $\phi$ about $v=175\gev$ 
in terms of the physical Higgs and Nambu-Goldstone bosons, 
\beq
\phi=\left(\begin{array}{c} G^+ \\ v+\frac{1}{\sqrt2}(h+iG^0)
\end{array} \right), 
\eeq 
yields $m_h^2=-2\mu^2=4\lambda v^2$ and all the Nambu-Goldstones
are massless. 

In scattering processes of gauge and Higgs bosons (at $\sqrt{s}\gg m_W$), 
the amplitudes can be decomposed into partial waves. Replacing the
longitudinal components of the gauge bosons with Nambu-Goldstone
bosons via the equivalence theorem~\cite{CG}, that decomposition takes
the form: 
\beq
a_J(s)=\frac{1}{32\pi}\int d(\cos\theta)\, P_J(\cos\theta)
\CM(s,\theta)
\eeq
for $a_J$ the $J$-th partial wave, $P_J$ the $J$-th Legendre polynomial
and $\CM$ the scattering matrix element. 
Recalling the scattering unitarity circle, one immediately notes that
while $|\im a_J|\leq 1$, the constraint on the real part is stronger:
\beq
\left|\re a_J\right| \leq \frac12.
\eeq
Now, in the absence of a physical Higgs boson,
various Nambu-Goldstone scattering amplitudes grow with energy so that
their tree-level amplitudes eventually exceed the unitarity bound.
For example, the $J=0$ amplitude for $W^{+}_LW^{-}_L\to Z_LZ_L$ scattering is
simply
\beq
a_{J=0}(s)=\frac{s}{32\pi v^2}-\frac{s}{32\pi v^2}\frac{s}{s-\mh^2}
+\CO\left(\frac{m_W^2}{s}\right)\stackrel{s\gg m_W}{\verylongrightarrow}
\frac{\mh^2}{32\pi v^2}
\eeq
where the second term is due to Higgs exchange and cuts off the bad high
energy behavior of the first term. The most divergent scattering
amplitude is for the mixed 
$2W^{+}_LW^{-}_L+Z_LZ_L$ ($I=0$) channel. Here~\cite{unitarity}
\beq
a_{J=0}\longrightarrow -\frac{5\mh^2}{64\pi v^2},
\eeq
which is already real. Bounding this by $\frac12$ yields 
\beq
\mh < 780\gev.
\eeq
Without a Higgs, the amplitude grows as $a_{J=0}=s/32\pi v^2$, so that 
the theory violates the unitarity bound at $\Lambda_U=1.2\tev$; that
is, either new physics must appear or the theory will become strongly
coupled at $\Lambda_U$. The bound on $\Lambda_U$ is slightly weaker
than that on $\mh$ only because it is more general; the Higgs
cancellation is the special case in which the new physics is the Higgs 
itself.

So, in summary, unitarity tells us that either ({\it i}\/) the Higgs
is lighter than about $780\gev$, or ({\it ii}\/) new physics must
appear below $\Lambda_U=1.2\tev$.

\subsection{Triviality}

The second classic constraint, triviality~\cite{trivial}, 
derives its name from the observation in $\varphi^4$-theory that the
quartic self-coupling $\lambda$ monotically increases as a function of the
momentum scale $Q$, becoming non-perturbative at the so-called Landau
pole, $\Lambda_T$.
Thus any construction of $\varphi^4$-theory which is to remain perturbative 
at all scales must have $\lambda=0$ identically, thus rendering the theory
``trivial.'' A similar observation can be made for the $\lambda$-coupling in
the Higgs sector of the standard model, 
with one important exception: the renormalization
group flow of $\lambda$ receives contributions with both signs. Namely, 
\beq
16\pi^2\frac{d\lambda}{d\log Q}=24\lambda^2 - (3g'^2+9g^2-12y_t^2)\lambda
+\frac38 g'^4+\frac34 g'^2g^2+\frac98g^4-6y_t^4+\cdots
\label{rge}
\eeq
where $y_t$ is the top quark Yukawa coupling and
the ellipsis represents contributions from small Yukawas
and higher loops. The first term
on the right side leads to the inflation of $\lambda$ in the 
ultraviolet, while the last term dominates contributions 
which drives $\lambda$ to asymptotic
freedom. Once again, given $\mh$, all terms in the renormalization group
equation (RGE) are known and this
equation can be solved simultaneously with the RGEs for $g$, $g'$ and $y_t$.
For large Higgs masses, the first term in Eq.~(\ref{rge}) dominates so
that
\beq
\Lambda_T\simeq Q \exp\left[\frac{2\pi^2}{3\lambda(Q)}\right]
\eeq
for any $Q<\Lambda_T$.
Larger Higgs masses obviously lead to Landau poles at lower scales. Thus for
any value of $\Lambda_T$, there is a corresponding maximum value of $m_h$ for
which the theory remains perturbative at all scales below the cutoff.

\begin{figure}[t]
\centering
\epsfxsize=4truein
\hspace*{0in}
\epsffile{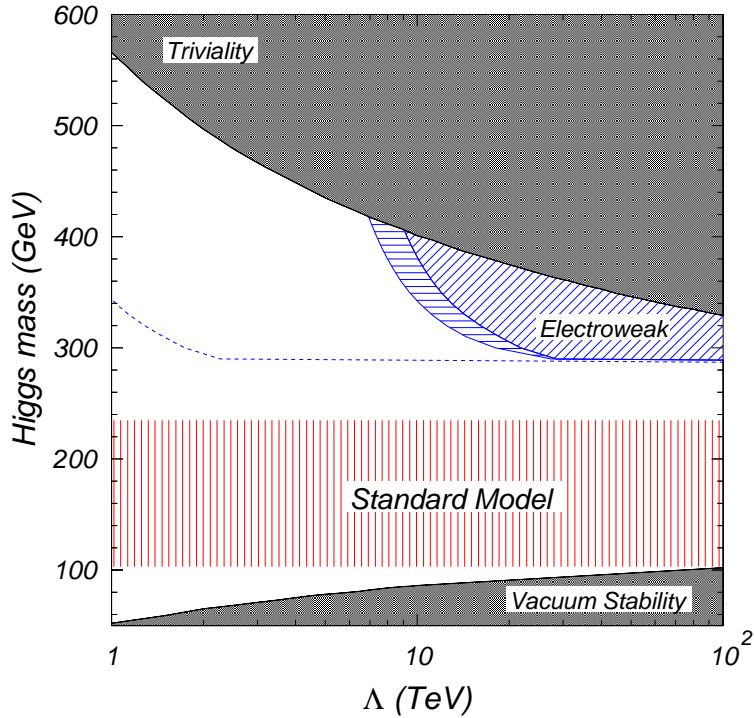}
\caption{The classic constraints on the $m_h$ -- $\Lambda$ plane,
  including triviality (dark region at top) and vacuum stability (dark
  region at bottom). The hatched regions marked ``Electroweak''
  and the region bounded by the dashed line are ruled out by precision 
  electroweak analyses; see the text for details. The region marked
  ``Standard Model'' is {\em allowed}\/ within the standard model, 
  bounded by the LEP mass bound below and the
  95\% CL electroweak precision bound above.}
\label{classic}
\end{figure}

In Figure~\ref{classic}, we show the maximum value of 
$\Lambda_T$ consistent with a given
$m_h$, calculated using the full coupled 2-loop RGEs 
(upper shaded region is excluded). But how robust is this bound? 
Once we allow new physics at a scale $\Lambda$
into the picture, it is clear
that there may appear additional operators that can contribute to the
physical Higgs mass without similarly contributing to its 4-point
interaction. Such new physics could increase or decrease
the triviality constraint on $\mh$, or even eliminate it
altogether~\cite{chanowitz}. 
Furthermore, $\Lambda$ and $\Lambda_T$ need not coincide exactly.

These considerations lead to two conclusions:
first, the triviality bound cannot be considered some sort of
model-independent upper bound on $\mh$ --- it is merely a bound above
which new physics near the Landau pole is implied. Second, the scale
of that new physics, while being near the Landau pole, could be a few
times larger. This last lesson is one that will be true throughout
most of these discussions: the exact scale at which some new particle
is produced can differ from the scale $\Lambda_T$ implied by triviality,
vacuum stability or precision electroweak physics by ${\cal O}(1)$
coefficients. The good news is that one generally expects the new
particles to be {\em lighter}\/ than $\Lambda_T$ if the new
physics is perturbative (consider the example of the $W$ mass versus
$1/\sqrt{G_F}$).
%%%%%%%%%%%%%%%%%%%%%%%%%%%%%%%%%%%%%%%%%%%%%%%%%%%%%%%%%%%%%%%%%%%
%%%%%%%%%%%%%%%%%%%%%%%%%%%%%%%%%%%%%%%%%%%%%%%%%%%%%%%%%%%%%%%%%%%

\subsection{Vacuum Stability}

The third and final classic constraint, vacuum stability~\cite{stability}, 
can also be seen as coming (at a very simplified level) from
the RGE for $\lambda$. Notice in Eq.~(\ref{rge}) that if $m_h$ is very {\it
small}\/ then the top Yukawa contribution dominates, driving $\lambda$ 
negative in the ultraviolet. At the turn-over scale, $\Lambda_V$, the Higgs
potential
becomes unbounded from below and the Higgs vev falls off to infinity (or at
least to $\phi\sim\Lambda_V$). Defining $\Lambda_V$ by $\lambda(\Lambda_V)=0$,
and keeping only the top quark Yukawa:
\beq
\Lambda_V\simeq Q \exp\left[\frac{8\pi^2\lambda(Q)}{3y_t^4(Q)}\right]
\eeq
for any $Q<\Lambda_V$.

A more complete and correct analysis requires the use of the
renormalization group-improved effective potential. Rather than
discussing the subtleties involved in such a calculation, we will
simply quote the recent results of Casas \etal~\cite{casas}. Note that
questions of tunnelling rates versus the lifetime of the universe have
not been included in this result, but do not change the conclusions of
the present paper at all.

One could also consider the effects coming from including higher
dimensional operators in the potential~\cite{datta} as we will do in
the next section.  These operators can have the effect of lifting the
potential at large scales (\eg, $V\sim \phi^6/\Lambda^2$), or
equivalently, shifting the value of $\lambda$ associated with a given
Higgs mass. Whether the stability bound is raised or lowered depends
sensitively on the sign of the new operators. To be conservative we
would like to choose these signs to maximize the region allowed for
the Higgs. But as we will see, the direct constraint coming from LEP
makes this an unnecessary exercise.

In Figure~\ref{classic}, we show the maximum value of
$\Lambda=\Lambda_V$ for a given value of $m_h$ (lower shaded region is
excluded); alternatively, the plot can be interpreted as giving the
minimum value of $m_h$ consistent with a given cutoff. Note that for
the scales $\Lambda_V$ of interest to us here (those within a few
orders of the weak scale), the current LEP bound of $\mh>102.6\gev$
$(\sqrt{s}=196\gev)$~\cite{lepmh} already supersedes the bounds coming
from vacuum stability arguments. And thus there is also nothing to be
gained by weakening the bound through the addition of higher
dimensional operators.

\section{Precision Electroweak Constraints}

While triviality and vacuum stability only provide bounds on the Higgs
mass as a function of the scale of new physics, a more direct
constraint arises from precision electroweak measurements. A global
fit of the data from the $Z$-pole experiments (ALEPH, DELPHI, L3,
OPAL, SLD), from $\nu-N$ deep inelastic scattering experiments (NuTeV,
CHARM, CCFR, CDHS), from low-energy $\nu-e$ scattering experiments
(CHARM~II), from atomic parity violation experiments (on cesium and
thallium), and from top quark production (CDF, D0) provides very
strong constraints on the range of $m_h$ under the assumption that
there is no physics other than that of the standard model near to the
weak scale. The fits to $m_h$ are constrained primarily by the oblique
corrections, and thus we can frame our discussion in terms of the
usual $S$ and $T$ parameters~\cite{st}.  (For an up-to-date discussion
of the Higgs mass and the oblique parameters, see~\cite{electroweak}.
We will set $U=0$ for this analysis since most types of new physics
contribute little if anything to $U$.)

Assuming no physics beyond the standard model, one 
finds~\cite{electroweak,global}:
\beq
m_h=\left(98^{+57}_{-38}\right)\gev
\eeq
or a one-sided limit of $m_h<235\gev$ at 95\% CL. What
happens once new physics is included? Because 
the standard model already works so well, any deviations from it can
be expressed using effective Lagrangian techniques, \ie, 
in terms of an expansion in inverse powers of the cutoff, $\Lambda$.
All independent dimension-6 operators which are consistent with the
symmetries and particle content of the standard model have been
catalogued~\cite{dim6} and their contributions to $S$ and $T$ have been
studied~\cite{htgr}. Two operators can produce shifts in $S$ and $T$ so as to
significantly alter the usual electroweak extraction of $m_h$:
\bea
\CO_{BW}=\frac{f_{BW}}{\Lambda^2} gg' (\phi^\dagger \tau^a
W^a_{\mu\nu} \phi) B^{\mu\nu} 
&\longrightarrow &\Delta S=
-16\pi\frac{v^2}{\Lambda^2}f_{BW}\\
\CO_{\phi}=\frac{f_{\phi}}{\Lambda^2} (\phi^\dagger D_\mu \phi)
(D^\mu \phi^\dagger \phi)
&\longrightarrow& \Delta T=-\frac{1}{\alpha}\frac{v^2}{\Lambda^2}
f_{\phi}
\eea
where $\alpha$ is the fine-structure constant and 
$f_{BW}, f_{\phi}$ are unknown coefficients. Because the
dependence of $S$ and $T$ on $\mh$ is primarily or exclusively logarithmic,
small changes in their experimentally preferred values result in exponentially
large shifts in $\mh$. Thus the operators $\CO_{BW}$ and $\CO_{\phi}$
can greatly affect the extraction of $\mh$ even for relatively large
$\Lambda$~\cite{hk}.

Of course, the correspondence between $\mh$ and $\Lambda$ is
intimately tied to the values of $f_{BW}$ and $f_{\phi}$. These
coefficients are in turn derived from the physics at $\Lambda$ which
we do not know. Therefore we will consider three possible scenarios
for each $\CO_i$ $(i=\phi,BW)$:
\begin{description}
\item[Loop-level:] Physics at $\Lambda$ is perturbative and $\CO_i$ is
  generated via loops. Therefore $|f_i|\leq\frac{1}{16\pi^2}$.
\item[Tree-level:] Physics at $\Lambda$ is perturbative but
  $\CO_{\phi}$ is generated at tree-level: $|f_{\phi}|\leq1$.
  $\CO_{BW}$ however cannot be generated at tree-level~\cite{arzt}, so
  $|f_{BW}|\leq\frac{1}{16\pi^2}$.
\item[Strong-coupling:] Physics at $\Lambda$ is strongly coupled and
  all or some of the fields in $\CO_i$ are composite. Naive
  dimensional analysis combined with custodial symmetry breaking tell
  us that $|f_i|\leq1$.
\end{description}
These three cases were chosen because they represent the three most
interesting limits of how new physics might affect $S$ and $T$: the
contributions to both are large$\sim 1$, small$\sim\frac{1}{16\pi^2}$,
or the contributions to $T$ and much larger than those to $S$ (such as
occurs in the simplest $Z-Z'$ mixing models). The missing fourth case
in which the contributions to $S$ dominate over those to $T$ occurs in
only a few examples of which we know and usually (if not always) $S$
has a sign which is inconsistent with the data.

This last case requires some comment (see \cite{chivukula} for a more
complete discussion). If the Higgs field itself is a composite, then
naive dimensional analysis~\cite{georgi} tells us that $f_{BW}\sim1$
while $f_{\phi}\sim16\pi^2$. On the other hand, those same arguments
would lead one to conclude that the top quark Yukawa $y_t\sim 4\pi$.
To avoid this (and other disasters), strongly coupled theories near
the weak scale should approximately conserve custodial isospin, $I$,
since the top quark mass is a $\Delta I=\frac12$ operator. Since
experimentally $y_t\simeq1$, the suppression associated with a $\Delta
I=\frac12$ insertion must be $\sim \frac{1}{4\pi}$. The operator
$\CO_{\phi}$ also breaks custodial isospin, but by $\Delta I=1$.
Creating it requires two $\Delta I=\frac12$ insertions, so we conclude
that it will naturally be suppressed by $\frac{1}{16\pi^2}$ over its
naive value. Thus $|f_{\phi}|\sim1$.

Arguments can be made for other choices of the coefficients, however
it seems to us that these choices cover the region of interest. Of
course, it is possible that different operators in the effective
Lagrangian can arise with very different coefficients. In fact, we are
counting on that since we are setting all operators other than
$\CO_{BW}$ and $\CO_\phi$ to zero, including flavor-violating
operators that would generally force the cutoff above 10 to $100\tev$.
If we included these operators at the same level as the two we are
considering, new physics would always be too weak to affect the
electroweak analysis. In fact, whenever one concentrates on new
physics at scales below 10 to 100$\tev$, one must demand (implicitly
or explicitly) that the new physics respects the flavor symmetries of
the standard model; the physics we are considering falls into this
class.

One could consider other operators which, while protecting the flavor
structure of the standard model, alter other aspects of the
electroweak analysis, such as the coupling of $Z$-bosons to electrons
or quarks~\cite{bs,xdim}. Barbieri and Strumia~\cite{bs} have
considered a ``statistical ensemble'' of models in which several
(flavor-conserving) operators were added to the standard model. They
found that over most of the phase space, non-universal corrections to
the standard model were so tightly constrained that the new
contributions to $S$ and $T$ were suppressed. While this is certainly
true, we are considering the more conservative questions of how large
$m_h$ could be, and how large $\Lambda$ could be consistent with a
given $m_h$. Adding more operators at the scale $\Lambda$ with fixed
coefficients will, over the great majority of parameter space, only
serve to rule out regions which we find to be allowed. It is possible
that by carefully tuning the other operators, one could actually allow
regions which we disallow, but this seems to us to be highly 
unlikely. Thus we feel that this analysis is as conservative as one
could reasonably demand.

However, to be conservative, we must not impose correlations on the
$f_i$ which may be unnatural.  Ideally, one would like to set only an
{\em upper bound}\/ on each operator and allow them to vary {\em
  independently}\/ within their allowed ranges so as to maximize
$\Lambda$ for a given $\mh$. So for example, the strong-coupling case
in which $|f_i|\leq1$ really does mean that any $f_i$ in the range
$-1\leq f_{\phi}\leq 1$ and $-1\leq f_{BW}\leq 1$ is allowed.  One
side-effect of this method is that there is no {\em lower}\/ bound on
$\Lambda$ as was found in \cite{hk}. However, it is easy to prove to
oneself that for the purposes of maximizing the $\mh$--$\Lambda$
correlation, one is always forced to one of the corners of this
rectangular parameter space where $f_{BW}=\pm1$ and $f_{\phi}=\pm1$
exactly. And because larger $\mh$ correspond to $S>0$ and $T<0$, we
already know that we will be using the corner in which
$f_{BW}=-f_{\phi}=+1$.

Analyses of this type have been completed by several
groups~\cite{hk,bs,others}. The procedure we follow here is the same
one that should be used once $\mh$ is measured. For each value of
$\mh$ between 50 and $1000\gev$, the full range of electroweak data is
fit to 2 unknown parameters, $S$ and $T$. A 95\% confidence region for
$S$ and $T$ is defined to be the region in which $\chi^2-\chi^2_{\rm
min} < 5.99$. $\Lambda$ is defined to be the maximum scale consistent
with the values of $S$ and $T$ at any point inside this region,
subject to the upper bounds placed on the $|f_i|$ by our choice of
scenario (loop-level, tree-level, or strong-coupling). More
specifically, for every point in the $(S,T)$ plane, there is a
corresponding set of $(\Lambda(S),\Lambda(T))$ which are the maximum
values of $\Lambda$ given the $f_i$ at their extremes derived from $S$
and $T$ respectively. Then for each point, we can define
$\Lambda(S,T)=\min(\Lambda(S),\Lambda(T))$. The overall $\Lambda$ is
then the maximum of the $\Lambda(S,T)$ within the 95\% confidence region.

One side-effect of this procedure which appears to be unavoidable is
that there exists a gap between the upper bound on $\mh$ in the
standard model, and the lower bound on the region in which new physics
can be excluded. Put another way, as $\Lambda\to\infty$, one would
expect that all values of $\mh>235\gev$ would be excluded, since
$235\gev$ represents the 95\% bound on $\mh$ in the standard model.
This is not the case for the following reason. In order to define
regions which are allowed and disallowed, one must calculate
$\chi^2_{\rm min}$ within the assumed model and then allow $\chi^2$ to
vary away from $\chi^2_{\rm min}$ by an amount consistent with the
number of degrees of freedom in the model. The standard model analysis
which generates a bound $\mh<235\gev$ produces a different
$\chi^2_{\rm min}$ than the procedure which sets $\mh=235\gev$ and
optimizes with respect to $S$ and $T$. Further, a 95\% {\em upper}\/
bound on $\mh$ is defined by setting $\Delta\chi^2=2.71$, whereas the
95\% region in the $(S,T)$ plane is defined using $\Delta\chi^2=5.99$.
Thus one does not expect the two analysis to have the same limiting
behaviors. Our conclusion is that, given current electroweak data, if
the Higgs mass falls in the region $235\gev\leq m_{h} \leq 290\gev$,
there will be clear evidence for the existence of new physics but no
statistically significant statement can be made 
about what that new physics scale may be.

Having explained our procedure, let us present our results. They are
summarized in Figure~\ref{classic}. The diagonally hatched area marked
``Electroweak'' represents the region ruled out for the
``strong-coupling'' class of models. The gap between the standard
model bound and the new physics bound is clearly visible. For
$\Lambda<10\tev$, the strong-coupling bound does no better than the
triviality bound; however, once $\Lambda>10\tev$, the effects of the
new operators become suppressed until finally at about
$\Lambda\simeq30\tev$, they are having no noticeable effect on the
fit.  The extension of this area which is horizontally hatched is the
{\em additional}\/ part of parameter space ruled out for the
``tree-level'' case.

Finally, the region above
the dashed line is ruled out for the ``loop-level'' case. Here we
notice that the new contributions to $S$ and $T$ are so small that
they have no noticeable affect on the fits for $\Lambda\gsim2.5\tev$,
and even for $\Lambda=1\tev$ one can derive a bound $\mh\lsim340\gev$.

We can draw any number of conclusions from these limits. First, it is
very hard to imagine that very weakly coupled physics
($f_i\sim\frac{1}{16\pi^2}$) is going to produce Higgs masses above
the standard model limit. Thus any discovery of $\mh\gsim 290\gev$ is
probably good evidence not only for new physics, but for physics with
$\CO(1)$ coefficients. Second, if $\mh\gsim 290\gev$, we can almost
certainly conclude that there must be new physics at scales less than
10 to $30\tev$, depending on how heavy that Higgs is. For example,
$\mh\gsim 400\gev$ requires new physics below $10\tev$.

\section{Fine-Tuning}

Discussions of fine-tuning inevitably invite acrimony over definitions
and degrees. There is a comfortable consensus that the mass parameter
of the standard model Higgs sector is unstable to ultraviolet
corrections and is thus fine-tuned. How to define that tuning is less
precisely discussed, in part because one usually considers the cutoff
to be far above the electroweak scale that precision in one's
definitions seems hardly worthwhile. But we are considering physics
very close to the weak scale, so we should take a moment to consider
our definition.

Going back to the original Higgs potential and decomposition into
physical and Nambu-Goldstone modes, notice that when we expand about some
general vacuum with $\phi=v$ undetermined, the resulting masses are
\bea
m_h^2&=&\mu^2+6\lambda v^2 \\ 
\mgb^2&=&\mu^2+2\lambda v^2.  
\eea 
where
$m_{GB}$ is the common mass of the Nambu-Goldstone bosons.  At the minimum
of the potential, $m_h^2=-2\mu^2=4\lambda v^2$ and all the Nambu-Goldstones
are massless. 

The issues we will be addressing concern the stability of the above
minimum, and in particular the parameter $\mu^2$ to radiative
corrections. And given that $m_W^2=g^2v^2/2$, it is clear that the
parameter $\mu^2$ is setting the electroweak scale, so we are really
discussing the stability of the electroweak scale.

We study the radiative corrections by considering the 
one-loop contributions to the effective potential, which have the form:
\bea
V_1(\phi)&=&\frac{1}{64\pi^2}\int d^4k\,
\str\log\left(k^2+M^2(\phi)\right) \\ 
&=& \frac{\Lambda^2}{32\pi^2}\str M^2(\phi)+\cdots 
\eea 
where the supertrace is defined by $\str=\mbox{Tr}\,(-1)^F$, and we are
keeping only the quadratically divergent pieces. This new contribution
can be absorbed into $V_0$ by shifting $\mu^2$: 
\beq
\mu^2\longrightarrow \bar\mu^2=\mu^2+\frac{\Lambda^2}{32\pi^2 v^2}\str
M^2(\phi) = \mu^2 + \frac{3\Lambda^2}{32\pi^2 v^2} (2 m_W^2 + m_Z^2 +
m_h^2 - 4m_t^2).
\label{str}
\eeq 
(In these expressions and those that follow, we are only keeping
the pieces of $\str M^2$ which are quadratic in $\phi$.\footnote{Naively, 
the supertrace is evaluated to be
$6m_W^2+3m_Z^2+m_h^2-12m_t^2$.  However, there is one subtlety that
must be recalled, namely that the trace is over the {\em
$\phi$-dependent}\/ masses, which for the scalar sector do not
correspond to the physical masses. In particular, 
$$
\sum_{scalars}M^2(\phi)=\left(m^2_h|_{\phi=v}-m^2_h|_{\phi=0}\right)+
3\left(\mgb^2|_{\phi=v}-\mgb^2|_{\phi=0}\right) 
=3m_h^2|_{\phi=v}.  
$$ 
Thus Eq.~(\protect\ref{str}) is reproduced.}) This last
expression is the source of the usual fine-tuning problem which arises
within the SM; if $\Lambda\gg m_W$, then we must suppose that the
tree-level $\mu^2$ and the loop contributions cancel to a very high
precision in order for $|\bar\mu^2|\sim m_W^2$. There is one
exception: Veltman famously 
noted~\cite{veltman} that there would be no (1-loop) fine-tuning
problem if only $\str M^2$ turned out to be zero. We will refer to the 
statement that $\str M^2=0$ (and its higher order generalizations)
as the ``Veltman condition.''

But while this condition $\str M^2=0$ is sufficient to cancel the 1-loop
quadratically divergent contributions to $V$, it is not an all-orders
result.  In particular, the general form of $\bar\mu^2$ is given
by~\cite{einhorn}: 
\beq 
\bar\mu^2=\mu^2+\Lambda^2\sum_{n=0}^\infty c_n(\lambda_i) \log^n (\Lambda/Q) 
\label{eq:mubar2}
\eeq 
where $c_0=(32\pi^2)^{-1}\str
M^2/v^2$ and the remaining $c_n$ can be calculated recursively via
\beq 
(n+1) c_{n+1}= \frac{dc_n}{d\log Q}=\beta_i \frac{\partial c_n}
{\partial \lambda_i},
\eeq 
from the requirement that the $\bar{\mu}^{2}$ should not depend on the 
renormalization scale $Q$.
For example, 
\bea
(16\pi^2)^2 c_1&=&\lambda\left(144\lambda-54g^2-18g'^2+72y_t^2\right)
+y_t^2\left(27g^2+17g'^2
+96g_3^2-90y_t^2\right)\nonumber \\
& &{}-\frac{15}{2}g^4+\frac{25}{2}g'^4+\frac92g^2g'^2.
\eea
Since each order in $n$ involves more
factors of $(16\pi^2)^{-1}$, the higher-loop contributions are
decreasingly important.

The usual approach to the higher-order Veltman condition is to demand
that each $c_n=0$ for all $n$ separately. In the infinite cutoff limit
(or if one wants a cutoff-independent solution), this would be the
correct procedure for solving the fine-tuning problem. And given that
the various $c_n$ are independent, the set of conditions $c_n=0$
vastly overconstrains the inputs ($\lambda_i$), so that no solution
exists. Thus one is usually led to conclude~\cite{einhorn} that the
Veltman condition is ultimately not useful for solving the fine-tuning
problem since it has no all-orders solution.

We would like to argue from a different limit.  For $\Lambda$ not much
larger than $m_W$, the stability of the weak scale does not require
such a dramatic set of cancellations.  This is easy to see. At 1-loop,
stability of the weak scale is only threatened if $\Lambda\gsim 4\pi
v\simeq 2\tev$.  Thus if new physics appears below about $2\tev$, it
poses no threat to the weak scale even if the Veltman condition is not
satisfied.  (We will make this argument slightly more sophisticated
shortly.) Now, were the 1-loop Veltman condition approximately
satisfied (presumably by accident), then $\Lambda$ could be pushed up
even higher. Two-loop quadratic divergences do not threaten the weak
scale until $\Lambda^2\log\Lambda\gsim (16\pi^2)^2 v^2$; that is,
$\Lambda\gsim 15\tev$. Thus if $\mh$ approximately satisfies the usual
Veltman condition, one could push new physics off to scales $\sim
15\tev$ without tuning.

This argument can be continued order by order. In particular,
three-loop divergences are irrelevant until $\Lambda\gsim 50\tev$.
Thus we could consider the possibility of scales up to $50\tev$ is the
2-loop Veltman condition were approximately satisfied. Since we are
only interested in scales of this order, we will work to 2-loop order
in the tuning condition.

This argument is slightly naive numerically, but it 
highlights the central truth that for cutoffs in the range that we might 
hope to probe any time in the foreseeable future: there is the
possibility of suppressing tunings through approximate solutions to
some finite-order version of the Veltman condition.
For any such scale, the absence
of large quadratic corrections is guaranteed simply by requiring
\beq
\sum_{n=0}^{n_{max}} c_n(\lambda_i)\log^n(\Lambda/m_{h}) =0
\label{eq:zero}
\eeq
for some $n_{max}$ chosen large enough so that the $(n_{max}+1)$-order 
corrections are automatically small. For the purposes of this paper, we will
set $n_{max}=1$; that is, we will go 1-loop beyond the lowest order Veltman 
condition. From the arguments above, this is a perfectly acceptable procedure 
for studying cutoffs below roughly $50\tev$. Note that there is no guarantee
that a solution to Eq.~(\ref{eq:zero}) even exists for large $n_{max}$; 
however we will see that for $n_{max}=1$ there is a solution. 

Since this is such an interesting possibility, we will consider in
more detail.
Because all the parameters in the condition $\str M^2=0$ have been measured
except for the Higgs mass, the Veltman condition predicts $m_h$. At LO, 
\beq 
m_h=(317\pm 11)\gev
\eeq
where the uncertainty is dominated by the measurement of $m_t$ which
we take to be $m_t=(174.3\pm 5.1)\gev$~\cite{Lancaster} where the
statistical and systematic errors are added in quadrature. At 
next-to-lowest order (NLO), the 
predicted Higgs mass decreases {\it and becomes cutoff-dependent!}\/ This 
should not surprise us since the presence of the logarithms in 
Eq.~(\ref{eq:zero}) prevents any one value of $m_h$ from satisfying
(\ref{eq:zero}) at all scales.
Another way to understand the $\Lambda$ dependence of the Veltman 
condition is the following:  Since the renormalized parameter 
$\bar{\mu}^{2}$ in Eq.~(\ref{eq:mubar2}) does not depend on the 
renormalization scale $Q$, we can in particular take $Q=\Lambda$; 
then all logarithms vanish except for $n=0$.  Therefore the Veltman 
condition is simply given by the lowest order expression 
$c_{0}(\lambda_{i}(\Lambda))=0$.  By rewriting the couplings 
$\lambda_{i}(\Lambda)$ in terms of the couplings which determine the Higgs 
mass $\lambda_{i}(m_{h})$, we again find Eq.~(\ref{eq:zero}) and we 
understand easily why the Veltman condition depends on the cutoff 
$\Lambda$.

However, this ``prediction'' assumes exact 
cancellation of the quadratic divergences, which is not actually necessary if
$\Lambda$ is close to the weak scale. This is another important difference 
that arises when one consider low cutoffs. Eq.~(\ref{eq:zero}) is more
correctly written as an inequality:
\beq
\sum_{n=0}^{n_{max}} c_n(\lambda_i)\log^n(\Lambda/m_{h}) 
< \frac{v^2}{\Lambda^2}.
\label{eq:finite}
\eeq
For $\Lambda=\mpl$, this implies a tuning of the Higgs mass (assuming any 
such value exists) to one part in $10^{16}$. Without a dynamical mechanism
for generating such a well-tuned value, we have simply replaced the 
fine-tuning between bare and loop-generated scalar masses with an equally bad 
fine-tuning among the various couplings of the theory. Though we may have a 
prediction of $m_h$, the fine-tuning 
associated with that value is unacceptable.
On the other hand, for low $\Lambda$, the tuning associated with any solution
to Eq.~(\ref{eq:finite}) is much smaller; 
we will discuss this in more detail shortly.

In this paper we will be interested in deriving $\Lambdamax$ as a
function of $m_h$. Clearly, as $m_h$ approaches $317\gev$, the value
of $\Lambdamax$ will increase rapidly until higher order corrections
cut it off; however we will find that $\Lambdamax$ can still be well
above the reach of current experiments for a wide range of $m_h$
centered around $317\gev$.

In order to continue, we must define more precisely what we mean by
fine-tuning.  Seemingly, the most natural definition would be the
sensitivity of the weak scale to the cutoff: $|\delta
m_W^2(\Lambda)/m_W^2|$, where $\delta m_W^2$ is the difference between
the tree and loop values, with all other quantities held fixed.  This
is the measure of fine-tuning used in \cite{BG}.  Varying the weak
scale as a function of $\Lambda$, we define
\begin{equation}
  \CF\equiv\left| \frac{\delta m_W^2}{m_W^2} \right|
  = \left| \frac{\delta v^2}{v^2} \right|
  = \left| \frac{\delta \mu^2}{\mu^2} \right|
  = \left| \frac{\delta m_h^2}{m_h^2} \right|
  = \frac{2\Lambda^2}{m_h^2}\left|\sum_n c_n \log^n(\Lambda/m_h)\right|.
%  = \frac{\Lambda^2}{16\pi^2v^2 m_h^2} \left| \str M^2 \right|.
\end{equation}
%where the last equality holds only at LO.
For any particular value of $\CF$, the weak scale is fine-tuned to one
part in $\CF$.\footnote{We consider the square of the weak scale to be
  the fundamental quantity to be set by ultraviolet physics rather
  than the weak scale itself.} In particular, $\CF\leq1$ represents
the absence of tuning. Note that a light Higgs corresponds to small
$\mu$ and thus greater fine-tuning as $\Lambda$ increases.

\begin{figure}[t]
\centering
\epsfxsize=4truein
\hspace*{0in}
\epsffile{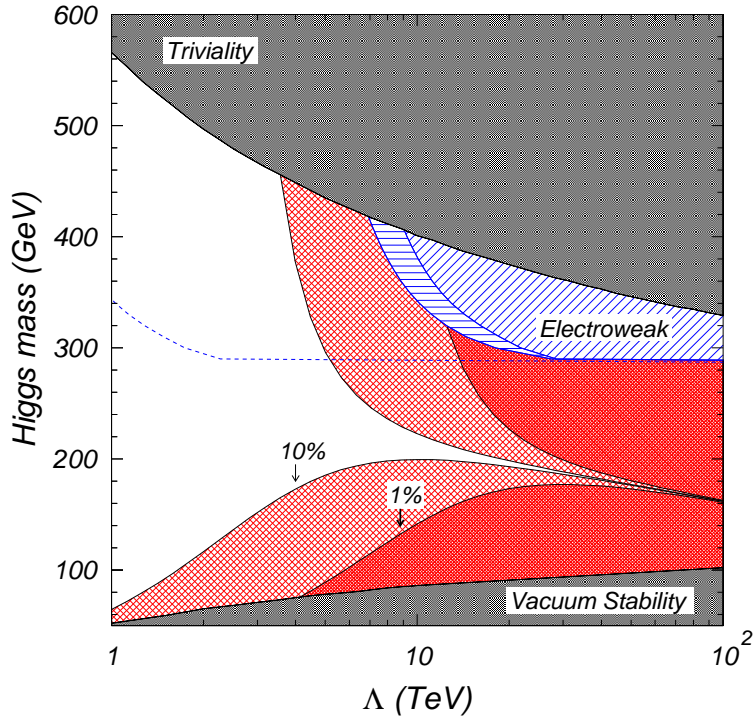}
\caption{Plot in the $\mh$ -- $\Lambda$ plane showing the canonical
  constraints from Figure~\protect\ref{classic} as well as the tuning
  contours. The darkly hatched region marked ``1\%'' represents
  tunings of greater than 1 part in 100; the ``10\%'' region means
  greater than 1 part in 10. The empty region is consistent with all
  constraints and has less than 1 part in 10 finetuning.}
\label{all}
\end{figure}

In Figure~\ref{all}, we plot $\Lambdamax$ as a function of $m_h$. The
hatched regions represent various values of the tuning parameter:
$\CF>10$ (light hatching) and 100 (dark hatching), ignoring
experimental uncertainties on $m_t$. Notice the light region extending
to very large $\Lambda$. This is the line along which the Veltman
condition is being approximately satisfied, and as expected, the value
$\mh$ which satisfies the condition is scale-dependent.

We conclude three things from staring at Figure~\ref{all}. First, if a
light Higgs is discovered, then new physics is very near at hand. For
example, for $\mh=130\gev$, finetuning of less than 1 part in 10
requires $\Lambda< 2.3\tev$. The reason for this is simple: smaller
$\mh$ means smaller $\mu$ which is all the harder to keep small. We
notice secondly that for a generic heavy value of $\mh$, new physics
must appear by about $3-5\tev$ in order to remain natural at the 10\%
level.  Finally, for values of $\mh$ in the range of $200\gev$, the
scale of new physics could turn out to be anomalously large, For
example, $\Lambda$ can be as high as $10\tev$ if $\mh$ falls in the
range $195\gev<\mh<215\gev$. We are not claiming that there exists a
dynamical reason why $\mh$ would fall in this range, but if it did,
our usual intuition about finetuning the electroweak scale would need
modification.

%%%%%%%%%%%%%%%%%%%%%%%%%%%%%%%%%%%%%%%%%%%%%%%%%%%%%%%%%%%%%%%%%%%
%%%%%%%%%%%%%%%%%%%%%%%%%%%%%%%%%%%%%%%%%%%%%%%%%%%%%%%%%%%%%%%%%%%

\section{The Supersymmetric Case}

After arguing that the scale of some general form of new physics could
naturally sit at several TeV, it is useful to consider a specific
example, namely that of supersymmetry (SUSY). Usually when one
considers SUSY models, one is led to conclude that fine-tunings of 1
part in 100 require the SUSY particle spectrum to lie below about
$1\tev$~\cite{BG,susyft}. 
Kane and King~\cite{kk} have even argued for gluinos lighter
than about $500\gev$ in some SUSY models in order to avoid tunings
greater than 1 part in 10.  Why is it, given the fact that SUSY is
supposed to stabilize the weak scale by eliminating quadratic
divergences, that the cutoff associated with SUSY so much smaller than
those we have previously quoted?

The short answer is that the previous limits were for a general form
of new physics; specific types may require much lower scales to avoid
tunings. The long answer is that it {\em is}\/ possible to arrange the
parameters of SUSY so as to allow multi-TeV spectra without tunings,
however in most models this behavior is not realized.

In SUSY models, one usually defines fine-tuning starting from the same
point as above, namely the minimization of the Higgs potential. In
SUSY, the {\em tree-level}\/ minimization already depends explicitly
on several masses in the problem. Specifically, one finds:
\beq
\frac12 m_Z^2 = \frac{m_{H_1}^2-m_{H_2}^2\tan^2\beta}{\tan^2\beta-1}
-\mu^2
\eeq
where $H_1$ and $H_2$ are the two Higgs fields, $\tan\beta$ is the
ratio of their vevs, and $m^2_{H_1}$, $m^2_{H_2}$ and $\mu$ are all
mass parameters. From this minimization condition, there are clearly
two ways to get equality: either all mass parameters on the righthand side
are $\sim m_W^2$, or they are much larger but cancellations occur so
that this particular function of their masses is $\sim m_W^2$. The first
possibility is considered to be natural, representing no tuning. The
second represents a tuning of roughly one part in
$m_Z^2/\max(m_{H_{1,2}}^2, \mu^2)$. Since most models produce
$m^2_{H_{1,2}}\sim\mu^2\sim \msusy^2$, this translates into
tunings of 
\beq
\CF\simeq m_Z^2/\msusy^2.
\eeq

Note the absence of any loop suppression factors in this expression.
It is this absence of additional $4\pi$ suppressions that makes SUSY
models seem more tuned at $1\tev$ than general models are at $5\tev$.

This comparison, however, is artificial. For the general models, we
{\it assume}\/ without argument that the tree-level mass term $\mu$
which sets the weak scale is naturally $\sim m_W$; we then only
consider the effect of new physics through loops. In the usual SUSY
case, we {\it derive}\/ the weak scale from the SUSY scale already at
tree-level, and so we are asking much more from our SUSY theory than
in the more general case.

It is possible to make a more fair comparison. We could dictate that
the mass parameters that enter the Higgs potential be weak-scale and
then study their sensitivity to loop corrections coming from the
remaining SUSY fields set at some higher mass scale $\msusy$. Because
of the special properties of SUSY, the original quadratic divergences
disappear, though they reappear in the guise of logarithmic
divergences. As an example, the log-divergent 1-loop correction to 
$m_{H_2}^2$ is given by:
\beq
\delta m_{H_2}^2 = \frac{1}{16\pi^2}\left(-6 g^2 M_2^2 -2 g'^2 M_1^2
+6 y_t^2 (m^2_{\tilde t_L}+m^2_{\tilde t_R}+m^2_{H_2}+A_t^2)
\right)\log\Lambda
\eeq
where the contributions from gauginos, top squarks and $A$-terms are
shown. Under most circumstances, this will be dominated by
$m_{\tilde t_L}\simeq m_{\tilde t_R}\equiv\msusy$ so that
\beq
\delta m_{H_2}^2\simeq \frac{3y_t^2}{4\pi^2}\msusy^2\log\Lambda.
\eeq
Except for the logarithm, this is exactly what we would expect for a
more general theory. In SUSY models, the role of the cutoff $\Lambda$
is played by the scale at which SUSY is broken in the standard model
sector. In the usual supergravity-type models, one takes $\Lambda
\sim 10^{16-18}\gev$, in which case the logarithm enhances the
correction by about 30. This 30 cancels the loop-suppression factor
and we again conclude that SUSY must sit below about $1\tev$.

To those familiar with SUSY, nothing mysterious has happened here. We
have simply restated the well-known fact that if we set the Higgs mass
parameters to be much smaller than $\msusy$ at the cutoff, then the
renormalization group flow down to the weak scale will tend to drag
them back to $\msusy$.

Several special cases have been discussed in the literature and we
only mention them here. It is possible to arrange the spectrum at the
cutoff scale so that the Higgs mass parameters are naturally driven to
be small at the weak scale~\cite{supernatural}, but such a scenario
requires tuning at the cutoff. We could alternatively choose to push a
subset of the SUSY spectrum to the weak scale, keeping the rest 
heavy~\cite{moreminimal}.  Obviously, 1-loop naturalness is guaranteed if
the top squarks and gauginos are $\sim m_W$. In such a case 2-loop
corrections~\cite{murayama} require that $\msusy\lsim 5-10\tev$ in
line with our 1-loop naturalness bounds in the general model.

Finally, there is a limit in which SUSY models look {\it exactly}\/
like the general case, but two requirements must be met: first, the
cutoff itself must be approximately $\msusy$ in order to kill the
logarithm; second, the underlying physics which sets the Higgs mass
parameters must set them to be $\sim m_W\ll\msusy$ at lowest order.
The first requirement is met, for example, in models with
gauge-mediated SUSY-breaking, but we do not know how to simultaneously
meet both. If both of these are met, loop corrections can be small and
the cutoff (and SUSY with it) can be safely moved into the multi-TeV
regime.

\section{Conclusions}

In this letter, we studied the theoretical constraints on the Higgs
mass and new physics scales in the minimal standard model.  In
addition to the constraints coming from unitarity, triviality,
vacuum stability and the electroweak precision analysis, we discussed
the constraint from the absence of fine-tuning in the Higgs mass
parameter (the Veltman condition).  As is well-known, the classic 
constraints allow a very high scale of new physics for a Higgs in the
mass range $100 \gev \lesssim m_{h} \lesssim 200 \gev$.
However, the fine-tuning condition places a significant constraint on
the new physics scale for this Higgs mass range.  For most Higss mass
values, we need new physics below 4 to 7~TeV (15 to 20~TeV) 
if we tolerate 10\% (1\%) fine-tuning.  
On the other hand, there is a narrow throat around
$m_{h} \sim 200$~GeV which could allow much higher new physics scale.
We noted, however, that a known specific example of new physics which
avoids fine-tuning naturally (supersymmetry) is constrained even more
tightly.  Even though one cannot draw a strict numerical limit from these
observations, the results presented here shed new light on the scale
of physics beyond the minimal standard model.

\section*{Acknowledgements}
We would like to thank M.~Chanowitz and L.~Hall for many interesting
discussions of the ideas presented here.  This work was supported in
part by the Department of Energy under contract DE--AC03--76SF00098
and in part by the National Science Foundation under grant
PHY-95-14797.

\end{document}